\documentclass[lettersize,journal]{IEEEtran}

\usepackage{amsmath}
\usepackage{cite}
\usepackage{color}
\usepackage[margin=15mm]{geometry}
\usepackage{graphicx}
\usepackage{hyperref}
\usepackage[separate-uncertainty]{siunitx}

\begin{document}

\title{Cross-correlation on a single channel for resistance noise measurements}
\author{
    Tim Thyzel%
    \thanks{Tim Thyzel is a staff scientist at Physikalisches Institut, Goethe-Universität Frankfurt, Frankfurt am Main, Germany}
}

\maketitle

\begin{abstract}
    Cross-correlation is an established tool to reduce the background in resistance noise measurements.
    However, the conventional method requires the amplifier, demodulator and digitizer channels to be duplicated, increasing the cost and complexity of the measurement circuit.
    We propose an alternating-current technique that allows cross-correlation with only a single channel by modulating the device under test with two carrier frequencies simultaneously.
    Using multiple software-based demodulators, we show that this method produces accurate amplitude measurements and noise spectra.
    The signal-to-noise-ratio is improved by $\SI{7}{\deci\bel}$ for standard parameters.
    Longer measurement durations increase this improvement, which makes the new technique a true cross-correlation method.
\end{abstract}


\DeclareSIUnit\voltrms{Vrms}

\section{Introduction}

The objective of resistance noise measurements is to measure small fluctuations in the resistance of a device under test (DUT) in the frequency range from Millihertz to Kilohertz \cite{Thyzel2024Methods}.
This can be challenging because measurement instrumentation usually exhibits $1/f$ noise (also called ``flicker`` or ``pink noise``) in the same frequency range.
Although modern semiconductor amplifiers achieve input noise spectral densities of the order $\SI[per-mode=symbol]{4}{\nano\volt\per\sqrt\hertz}$ at $f = \SI{1}{\hertz}$, this can already be sufficiently high to mask the fluctuations of a very ``quiet`` DUT whose excitation power is thermally limited.

To overcome the problem of low-frequency amplifier background noise, two methods are usually employed:
Amplitude modulation with a sinusoidal excitation current shifts the spectrum of the DUT noise to higher frequencies, where the amplifier noise is lower \cite{Scofield1987}.
This circumvents the low-frequency $1/f$ noise, but white noise originating from the amplifier remains present at all frequencies.
To suppress the white noise contribution as well, a cross-correlation technique with multiple parallel amplifiers needs to be used \cite{Sampietro2000}.
Since the random fluctuations produced by the semiconductor components in the amplifier circuits are mutually independent if the circuits are isolated from each other, they are not correlated.
Therefore, calculating the cross-correlation function, or its equivalent in frequency space, between the outputs of physically distinct amplifier devices suppresses the noise background and can reveal even lower DUT noise levels.

While the cross-correlation technique is well-established \cite{Sampietro1999, Sampietro2000, Rubiola2010, Achtenberg2021}, the need for multiple parallel amplifier chains may be problematic with regards to setup cost.
Amplifiers suitable for resistance noise measurements are available commercially only at great expense, or must be custom-designed by experts \cite{Scandurra2022} to fulfill the requirements for extremely low input noise, interference rejection and shielding, power supply noise rejection, programmable gain, and electrical safety.
Additionally, each amplifier requires its own demodulator, often in the form of a discrete audio-frequency lock-in amplifier, and its own simultaneously-sampling analogue-to-digital converter (ADC, also called ``digitizer``) .
Consequently, the cost of a two-channel cross-correlation circuit may be almost twice that of the corresponding single-channel setup.

In this work, we propose a cross-correlation method using only a single amplifier.
Instead of relying on the fact that the noise of distinct instruments is uncorrelated, we exploit the lack of correlation between two non-overlapping frequency bands in the white background noise of the same instrument.
We demonstrate that, by using two superimposed sinusoidal excitation currents and two software-based demodulators, it is possible to perform cross-correlation noise measurements without the need for parallel amplifiers.
After describing the implementation, we will verify that the new technique reproduces the DUT noise accurately and that the signal-to-noise ratio is improved by cross-correlation as expected.

At this point, the scope of the proposed method should be clearly defined:
First and foremost, it is applicable to resistance noise measurements, in which the signal to be measured can be modulated.
Measurements of the input-referred voltage noise of amplifiers, for instance, do not fall into this category.
Second, it must be possible to excite the DUT using an alternating current without introducing harmonic distortion.
This condition is not fulfilled if the device impedance is strongly current-dependent, as e.g.\ in diodes.
Third, the device impedance must remain constant over the range of reference frequencies used for modulation.
In summary, the method is well suited for the purpose of material characterization, e.g.\ in semiconductor defect spectroscopy \cite{Fleetwood2015}, evaluation of resistive temperature \cite{Ryger2017} or strain sensors \cite{Schwalb2010}, or fundamental condensed matter research \cite{Müller2025}.

\IEEEpubidadjcol  

\section{Implementation}

\begin{figure*}
    \centering
    \includegraphics[scale=1.1]{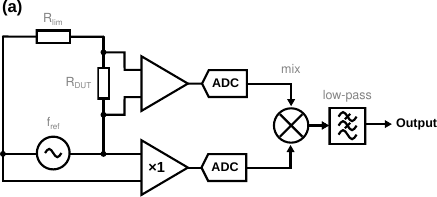}
    \hspace{2mm}
    \includegraphics[scale=1.1]{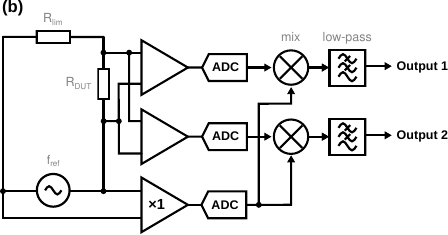} \\
    \vspace{5mm}
    \includegraphics[scale=1.1]{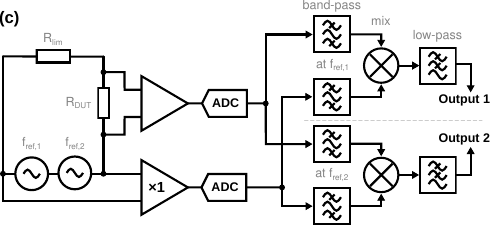}
    \caption{\textit{(a)} Conventional single-channel, single-reference lock-in demodulator circuit.
        A sinusoidal current is passed through $R_\mathrm{DUT}$ and its voltage drop is amplified, recorded, and demodulated by the reference sine wave through mixing and low-pass filtering.
        The ``x1`` symbol indicates that the reference voltage usually does not need to be amplified, only converted from a differential/floating to a single-ended/ground-referenced signal.
        \textit{(b)} Conventional dual-channel circuit for cross-correlation, which requires two parallel amplifier-ADC channels sampling the DUT voltage.
        \textit{(c)} New single-channel, multi-reference circuit with only one amplifier-ADC channel but two superimposed reference signals and two corresponding demodulator units.
    }
    \label{fig:circuits}
\end{figure*}

We implement the new ``multi-reference`` method by extending the standard lock-in demodulator circuit shown in Fig.\ \ref{fig:circuits}a.
For this conventional ``single-reference`` setup, the device-under-test (DUT) resistance $R_\mathrm{DUT}$ is excited with a sine voltage at the reference (or carrier) frequency $f_\mathrm{ref}$ from a function generator (Stanford Research Systems DS360).
A series resistance $R_\mathrm{lim}$ limits the current through the DUT.
The voltage drop over the DUT, formed by $R_\mathrm{DUT}$ modulated by the sine current, is amplified and then recorded by an analogue-to-digital converter (ADC), in this instance one of the four simultaneously-sampled channels of an NI (formerly National Instruments) 9239 data acquisition unit.
Standard lock-in demodulation then takes place in the digital domain by mixing and low-pass filtering with the reference voltage recorded on another ADC channel, yielding an output signal proportional to $R_\mathrm{DUT}$.
Note that, although discrete high-gain amplifiers are normally used, the built-in differential buffer amplifier of NI 9239 was sufficient for this demonstration.

Usually, to enable cross-correlation measurements the circuit would be extended as shown in Fig.\ \ref{fig:circuits}b.
The amplifier and ADC recording the DUT voltage drop would need to be duplicated, and a second demodulator added, which shares the reference signal with the first demodulator.
The noise background of the first signal processing chain is uncorrelated with that of the second, so that a cross-correlation analysis of the two output signals would suppress the background, but at the cost of additional hardware.

The new method shown in Fig.\ \ref{fig:circuits}c does not require two amplifier channels for cross-correlation.
Instead, its hardware is almost identical to that of the single-channel circuit in Fig.\ \ref{fig:circuits}a. 
However, the function generator now outputs a superposition of two sinusoidal voltages -- $V_\mathrm{ref,1}$ at frequency $f_\mathrm{ref,1}$, $V_\mathrm{ref,2}$ at $f_\mathrm{ref,2}$ -- in its two-tone mode.
Consequently, the spectrum of the DUT voltage drop contains two copies of the DUT resistance noise, each modulated by a different reference frequency.
After amplification and recording both the DUT voltage drop and the reference signal are passed into two parallel demodulator stages, one for each reference frequency $f_\mathrm{ref,i} : i \in \{1,2\}$.
Each stage begins by extracting a frequency band around $f_\mathrm{ref,i}$ in both the DUT and the reference signal using two identical band-pass filters.
The infinite-impulse-response filters are of the Butterworth type, chosen for its flat pass band important for noise spectroscopy.
The filter bandwidth is chosen large enough to encompass the low-frequency DUT resistance noise spectrum, whose measurement is the objective.
After mixing and low-pass filtering, each demodulator outputs a signal proportional to $R_\mathrm{DUT}$ and $V_\mathrm{ref,i}$.
The demodulators are implemented in a Python software package allowing live-streaming digital signal processing, whose correct function and good noise properties we have demonstrated in Ref.\ \cite{Thyzel2025SoftwareLockin}.
We make the package available under an open-source license at Ref.\ \cite{Thyzel2026SoftwarePackage}.
Thanks to this software-based demodulator solution, duplicate audio-frequency lock-in amplifiers are not required, keeping the setup cost identical to the simple single-reference circuit in Fig.\ \ref{fig:circuits}a.

The output signals of both demodulators still contain the amplifier background noise, but, crucially, the backgrounds should be uncorrelated since they originate from different frequency bands centred at $f_\mathrm{ref,i}$.
This fact can now be exploited to suppress the amplifier noise and improve the signal-to-noise ratio.

\section{Performance tests}

\subsection{Selection of DUT and reference frequencies}\label{sec:selection-of-dut-and-reference-frequencies}

For this demonstration, we chose a crystal of the organic molecular metal $\theta$-(BEDT-TTF)\textsubscript{2}-CsCo(SCN)\textsubscript{4} as the device under test (DUT), because its resistance is known to exhibit strong $1/f$ noise \cite{Thomas2022}.
Since we merely use the DUT as a generator for $1/f$ resistance noise, the details of the material and the crystal are not a subject of this work.

\begin{figure}
    \centering
    \includegraphics[scale=1.0]{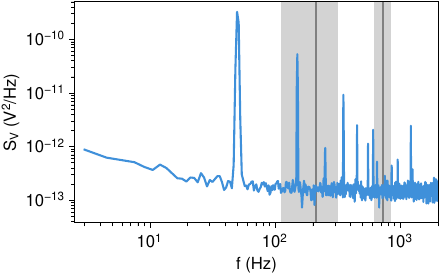} \\
    \vspace{2mm}
    \includegraphics[scale=1.0]{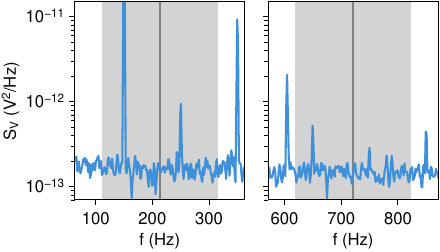}
    \caption{\textit{Top:} Background noise in the multiple-reference circuit.
        Vertical grey lines mark the reference frequencies, light-grey shaded areas show the pass bands that will be contained in the demodulated signals later.
        \textit{Bottom:} Same data as top, zoomed in on the reference frequencies.
    }
    \label{fig:background-without-modulation}
\end{figure}
Next, we selected reference frequencies $f_\mathrm{ref}$ according to the following criteria:
$f_\mathrm{ref}$ must not coincide with any interference peak in the power spectrum, because this would give a spurious contribution to the DC value of the demodulator output.
Also, the pass bands around the reference frequencies, which will be included in the demodulated output signals, should not overlap to avoid introducing correlation in the background noise between output signals.
To fulfill these criteria, we measured the noise background in the multi-reference circuit by setting $V_\mathrm{ref,1} = V_\mathrm{ref,2} = 0$ and recording the voltage drop $V(t)$ across the DUT without any lock-in demodulation.
The resulting power spectrum shown in Fig.\ \ref{fig:background-without-modulation} is dominated by the background noise of the amplifier and ADC, whose open-circuit input-referred white noise level of $S_V \approx \SI[per-mode=symbol]{2e-13}{\square\volt\per\hertz}$ was already known to us beforehand.  
The two reference frequencies we selected, $f_\mathrm{ref,1} = \SI{213}{\hertz}$ and $f_\mathrm{ref,2} = \SI{721}{\hertz}$, are marked by vertical lines in the figure and avoid the interference peaks at harmonics of $\SI{50}{\hertz}$, the European electricity grid frequency.
The pass-bands, marked by shaded areas and each $\SI{200}{\hertz}$ wide, do not overlap and mainly contain the white amplifier noise which we will later suppress by cross-correlation.
Note that the interference peaks that lie inside the pass-bands are very sharp and therefore unproblematic for broad-band noise spectroscopy.

\subsection{Verification of frequency response}

Next, we ensured that the multi-reference method reproduces the measurement results obtained by the conventional single-reference lock-in technique.
For this purpose, the same RMS current was applied to the DUT in both circuits:
In the single-reference circuit (see Fig.\ \ref{fig:circuits}a), the function generator output a sine wave at $f_\mathrm{ref} = \SI{213}{\hertz}$ with $V_\mathrm{ref} = \SI{3}{\voltrms}$.
In the multi-reference circuit (see Fig.\ \ref{fig:circuits}c), one function generator output at $f_\mathrm{ref,1} = \SI{213}{\hertz}$ and the other at $f_\mathrm{ref,2} = \SI{721}{\hertz}$, both with an amplitude of $V_\mathrm{ref,1} = V_\mathrm{ref,2} = \SI{3}{\voltrms} / \sqrt{2}$. 
This reduced amplitude ensures that the same power is dissipated in the DUT in both circuits.

\begin{figure}
    \centering
    \includegraphics[scale=1.0]{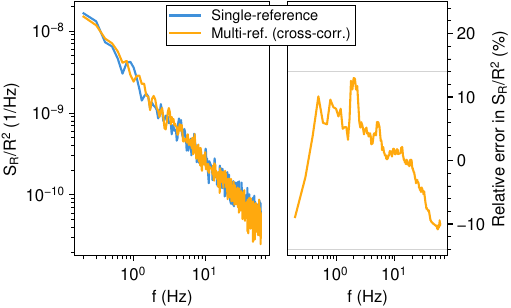}
    \caption{\textit{Left:} Good agreement of the normalized resistance noise power spectra between the single-reference and the multi-reference methods.
        Interference peaks have been removed in this data set.
        The multi-reference spectrum is the cross-PSD of the two output signals.
        \textit{Right:} When the spectra on the left are smoothed (linear Savitzky-Golay filter, half a frequency decade wide), the error in the multi-reference spectrum relative to the single-reference spectrum is below $\SI{14}{\percent}$ (horizontal bars).
    }
    \label{fig:frequency-response}
\end{figure}

The DUT resistance measured in the single-reference configuration is $R_\mathrm{DUT} = \SI{5.990}{\kilo\ohm}$, and the two measured values in the multi-reference setup -- demodulated at $f_\mathrm{ref,1}$ and $f_\mathrm{ref,2}$, respectively -- differ by less than $\SI{0.3}{\percent}$ from this.
In addition to these DC values, the DUT noise spectra $S_R(f)/R^2$ shown in Fig.\ \ref{fig:frequency-response} match that of the conventional single-reference method quite well:
The difference in power spectral density between the two methods does not exceed $\SI{14}{\percent}$, which is usually sufficiently accurate when the $1/f$ noise parameters of the DUT are investigated.
Note, however, that in some cases slight deviations from the $1/f$ behaviour, which may intrinsically occur in the DUT noise \cite[Fig.\ 2A/B]{Sasaki2017}, could be masked by this small measurement error.
In summary, the new multi-reference method measures both the DC value and the noise power spectral density with sufficient accuracy for most applications. 

\subsection{Signal-to-noise-ratio improvement}

\begin{figure*}
    \centering
    \includegraphics[scale=1.0]{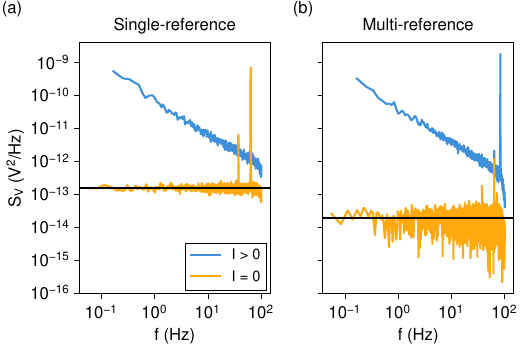}
    \includegraphics[scale=1.0]{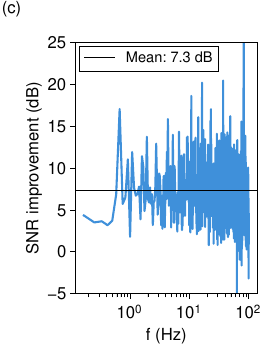}
    \includegraphics[scale=1.0]{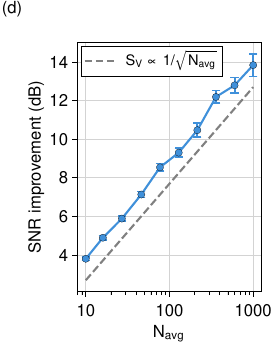}
    \caption{\textit{(a)/(b)}: Comparison of the voltage power density spectra for the conventional single-reference method (a) and the new multi-reference cross-correlation method (b).
        The measurement at non-zero excitation current $I > 0$ reflects the $1/f$-like resistance noise of the DUT, while the spectrum at $I = 0$ represents the background of the measurement setup.
        The mean low-frequency background PSD $S_V(f \leq \SI{30}{\hertz})$, shown by black horizontal lines, is far lower in the multi-reference setup thanks to cross-correlation.
        \textit{(c)} Computing the signal-to-noise-ratio $\mathrm{SNR}(f)$ for (a) and for (b), and then subtracting the two for each frequency $f$ gives the SNR improvement spectrum shown here.
        In this case, averaging over $N_\mathrm{avg} = \num{50}$ sections gave a mean improvement using the new cross-correlation method of $\SI{7.3}{\deci\bel}$ across all frequencies below $\SI{30}{\hertz}$.
        \textit{(d)} Performing the analysis in (c) for varying amounts of averaging sections increases the improvement in SNR, because the background cross-PSD $S_\mathrm{V}(f, I=0)$ scales as $1 / \sqrt{N_\mathrm{avg}}$.
    }
    \label{fig:signal-to-noise}
\end{figure*}
Having shown the new multi-reference method to be correct and accurate, we now investigate its signal-to-noise-ratio.
Two DUT voltage noise measurements were taken with each method, one at zero excitation current $I = 0$ to determine the noise background, and one at $I > 0$ from which the DUT resistance noise can be obtained.
For $I = 0$, the function generator was disconnected from $R_\mathrm{lim}$, and both poles of $R_\mathrm{lim}$ connected to $R_\mathrm{DUT}$.
For both currents, the power spectral density (PSD) $S_V(f)$ of the DUT voltage drop was computed by Welch's periodogram analysis.
Since two output signals are available in the multi-reference method, one for each reference frequency, the complex cross-correlation spectrum was computed by Welch's method, and its absolute magnitude taken. 
The voltage spectra in Fig.\ \ref{fig:signal-to-noise}a/b show that the background PSD is much lower in the multi-reference than in the single-reference method.
This is the central result of this work.

In the context of noise measurements, the signal-to-noise-ratio (SNR) can be defined as the ratio between the power spectral density for $I > 0$ (the DUT noise) and that for $I = 0$ (the background noise):
\begin{align}
    \mathrm{SNR}(f) = \SI{10}{\deci\bel} \times \log_{10}\left(\frac{S_V(f, I > 0)}{S_V(f, I = 0)}\right) \label{eq:snr-definition}
\end{align}
Fig.\ \ref{fig:signal-to-noise}c shows that the SNR improves by about $\SI{7}{\deci\bel}$ when the multi-reference method is employed.
We stress that this improvement comes at no additional hardware complexity compared to the conventional single-reference setup.
Note that some signal amplitude is lost because the excitation power is distributed to two frequencies, but the strongly suppressed background more than offsets this.

Having gained the ability to perform cross-correlation measurements, we can further improve the SNR simply by measuring for a longer time.
Contrary to amplitude measurements, where noise can be reduced by averaging over longer times, the background in PSD measurements cannot be lowered in this way, because the PSD computation discards phase information by taking the square of the Fourier transform.
However, it \textit{is} possible to decrease the background PSD by cross-correlating two signals whose background is uncorrelated.
Fig.\ \ref{fig:signal-to-noise}d shows that our new method provides two signals with fully uncorrelated backgrounds:
Increasing the number $N_\mathrm{avg}$ of averaging sections (also known as ``segments``) in Welch's periodogram, i.e.\ measuring for a longer time, improves the SNR with the same square-root scaling known from the conventional cross-correlation technique that requires multiple amplifier channels \cite[Fig.\ 3c]{Thyzel2024Methods} \cite[Eq.\ 11]{Rubiola2010} \cite[Fig.\ 6]{Scandurra2013}.
This proves that the new single-channel multi-reference method is a true cross-correlation technique.

\subsection{Relation to conventional cross-correlation}\label{sec:relation-to-conventional-cross-correlation}

\begin{figure}
    \centering
    \includegraphics[scale=1.0]{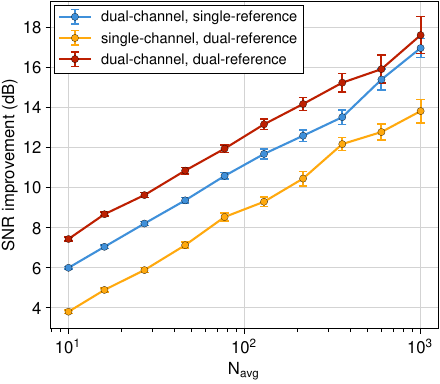}
    \caption{The same analysis as in Fig.\ \ref{fig:signal-to-noise} done for three different cross-correlation methods.
        The SNR improvement is relative to a single-channel, single-reference measurement.
    }
    \label{fig:methods-snr-comparison}
\end{figure}

To investigate how the new multi-reference cross-correlation method compares to the conventional multi-channel technique, we added a second amplifier and ADC recording the DUT voltage drop to the circuits.
The SNR data of this setup, shown in Fig.\ \ref{fig:methods-snr-comparison}, provides two insights:

First, the new method described in the previous sections (``single-channel, dual-reference``) is equivalent to the conventional cross-correlation method (``dual-channel, single-reference``) in the sense that the SNR improves at the same rate with longer measurement duration.
The cost of not having a second channel available in the new method is an inferiority in SNR of about $\SI{2}{\deci\bel}$, because excitation power is distributed to both reference frequencies to keep the power dissipated in the DUT constant, limiting self-heating.

Second, if a second channel \textit{is} available, the SNR can be improved further still by employing the new technique.
In the ``dual-channel, dual-reference`` configuration, one demodulated output signal is available from each amplifier for each frequency.
The cross-PSD is computed for each pair-wise combination of these four output signals, and an average is taken over these combinations in addition to section-averaging.
This procedure, analogous to cross-correlation with more than two amplifier channels \cite{Scandurra2013}, increases the SNR by a margin of $\SI{1.5}{\deci\bel}$ over the ``gold-standard`` dual-channel technique. 

\subsection{Extension to more cross-correlation pairs}

A logical extension to the method presented above would be the use of an arbitrary function generator, with which not only two, but a great number of reference/carrier frequencies can be generated.
Utilizing the full bandwidth of the DUT, amplifier and digitizer, this would allow for a large-scale cross-correlation that has until now required an array of many amplifiers \cite{Scandurra2013}.

It has to be noted, however, that the increase in signal-to-noise ratio, expected due to the large number of cross-correlation pairs, may saturate at a certain number of reference frequencies, because the ``signal amplitude`` is distributed to all those frequencies to keep the DUT power dissipation constant.
To estimate when this saturation takes place, we calculate how the signal to noise ratio $\mathrm{SNR}$ defined in Eq.\ \eqref{eq:snr-definition} depends on the number $N_c$ of reference frequencies.
From Eq.\ 14 in Ref.\ \cite{Scandurra2013}, we determine that the background noise PSD decreases with the number of unique reference frequency pairs according to a square-root relationship:
\begin{align}
    S_V(f, I = 0) \propto \frac{1}{\sqrt{\frac{N_c (N_c - 1)}{2}}}
\end{align}
To keep the DUT power $P = R_\mathrm{DUT} I^2$ constant, the excitation current $I$ must be decreased to $I / N_c$.
Since the DUT voltage drop PSD depends on the square current due to Ohm's law, it will decrease by the same factor:
\begin{align}
    S_V(f, I > 0) = S_R(f) \, \frac{I^2}{N_c^2} \propto \frac{1}{N_c^2}
\end{align}
The signal-to-noise-ratio is then:
\begin{align}
    \mathrm{SNR}(f) &= \SI{10}{\deci\bel} \times \log_{10}\left(\frac{S_V(f, I > 0)}{S_V(f, I = 0)}\right) \\
    &= \SI{10}{\deci\bel} \times \log_{10}\left(\frac{\sqrt{N_c (N_c - 1)}}{N_c^2}\right) + \mathrm{const.}
\end{align}%
\begin{figure}
    \centering
    \includegraphics[scale=1.0]{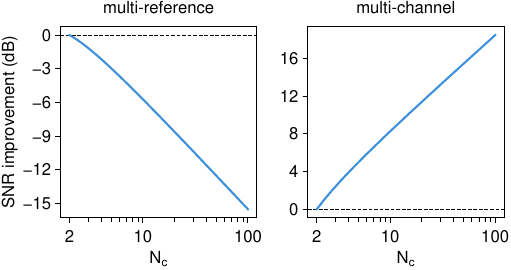}
    \caption{
        \textit{Left:} When the number $N_c$ of reference/carrier frequencies is increased in the single-channel-multi-reference setup, the signal-to-noise-ratio (SNR) does not improve but deteriorates.
        \textit{Right:} This is in contrast to the multi-channel-single-reference setup, where the SNR does improve when the number $N_c$ of channels is increased.
        In both figures, the SNR improvement is computed relative to $N_c = 2$.
    }
    \label{fig:massive-multichannel}
\end{figure}%
As shown in Fig.\ \ref{fig:massive-multichannel}, this function has a negative slope for $N_c \geq 2$, which means that the SNR does not improve any futher when the number of reference frequencies is increased above two.
This indicates that the optimal configuration is to use as many amplifier channels as possible in combination with exactly two reference frequencies to obtain the improvement by $\SI{1.5}{\deci\bel}$ demonstrated in Sec.\ \ref{sec:relation-to-conventional-cross-correlation}.

\section{Conclusion}

We have demonstrated a cross-correlation technique for resistance noise measurements which does not require multiple amplifiers.
Instead, multiple copies of the device-under-test (DUT) signal modulated by different carrier frequencies are recorded by a single amplifier-digitizer channel and then separated and demodulated using digital signal processing.
We have shown that this method accurately measures the frequency spectrum of the DUT noise and that, crucially, the amplifier background noise is uncorrelated between the spectrum copies.
By computing the cross power spectral density of the demodulated copies, we suppress the amplifier background and achieve an improvement in signal-to-noise ratio (SNR) by $\SI{7}{\deci\bel}$ for standard parameters.
Longer averaging times increase the SNR further, making this method a true cross-correlation technique.
As it requires no additional hardware compared to the setup without cross-correlation, the new method removes a barrier to the routine use of cross-correlation in resistance noise measurements, whenever alternating-current excitation is possible.

\bibliographystyle{IEEEtran}
\bibliography{paper}

@article{Thyzel2024Methods,
    doi = {10.1088/1361-6501/ad876f},
    url = {https://dx.doi.org/10.1088/1361-6501/ad876f},
    year = {2024},
    publisher = {IOP Publishing},
    volume = {36},
    number = {1},
    pages = {015501},
    author = {Tim Thyzel and Marvin Kopp and Jonathan Pieper and Tristan Stadler and Jens Müller},
    title = {Methods in fluctuation (noise) spectroscopy and continuous analysis for high-throughput measurements},
    journal = {Measurement Science and Technology},
}

@article{Scofield1987,
    author = {Scofield, John H.},
    title = "{ac method for measuring low-frequency resistance fluctuation spectra}",
    journal = {Review of Scientific Instruments},
    volume = {58},
    number = {6},
    pages = {985-993},
    year = {1987},
    month = {06},
    issn = {0034-6748},
    doi = {10.1063/1.1139587},
    url = {https://doi.org/10.1063/1.1139587},
}

@article{Sampietro1999,
    author = {Sampietro, M. and Fasoli, L. and Ferrari, G.},
    title = "{Spectrum analyzer with noise reduction by cross-correlation technique on two channels}",
    journal = {Review of Scientific Instruments},
    volume = {70},
    number = {5},
    pages = {2520-2525},
    year = {1999},
    doi = {10.1063/1.1149785},
    url = {https://doi.org/10.1063/1.1149785},
}

@misc{Rubiola2010,
      title={The cross-spectrum experimental method},
      author={Enrico Rubiola and François Vernotte},
      year={2010},
      eprint={1003.0113},
      archivePrefix={arXiv},
      primaryClass={physics.ins-det},
      url={https://arxiv.org/abs/1003.0113},
}

@ARTICLE{Fleetwood2015,
  author={Fleetwood, D. M.},
  journal={IEEE Transactions on Nuclear Science}, 
  title={$1/f$ Noise and Defects in Microelectronic Materials and Devices}, 
  year={2015},
  volume={62},
  number={4},
  pages={1462-1486},
  doi={10.1109/TNS.2015.2405852}
}

@article{Müller2025,
    author = {Jens Müller},
    title = {Noise and fluctuations in condensed matter systems},
    journal = {Contemporary Physics},
    volume = {66},
    number = {1-4},
    pages = {62--90},
    year = {2025},
    publisher = {Taylor \& Francis},
    doi = {10.1080/00107514.2025.2576319},
    URL = {https://doi.org/10.1080/00107514.2025.2576319},
}

@misc{Thyzel2025SoftwareLockin,
      title={Software-defined lock-in demodulator for low-frequency resistance noise measurements}, 
      author={Tim Thyzel},
      year={2025},
      eprint={2412.00093},
      archivePrefix={arXiv},
      primaryClass={physics.ins-det},
      url={https://arxiv.org/abs/2412.00093}, 
}

@article{Thomas2022,
  title = {Involvement of structural dynamics in charge-glass formation in strongly frustrated molecular metals},
  author = {Thomas, Tatjana and Saito, Yohei and Agarmani, Yassine and Thyzel, Tim and Lonsky, Martin and Hashimoto, Kenichiro and Sasaki, Takahiko and Lang, Michael and M\"uller, Jens},
  journal = {Phys. Rev. B},
  volume = {105},
  issue = {4},
  pages = {L041114},
  numpages = {6},
  year = {2022},
  month = {Jan},
  publisher = {American Physical Society},
  doi = {10.1103/PhysRevB.105.L041114},
  url = {https://link.aps.org/doi/10.1103/PhysRevB.105.L041114}
}

@article{Sasaki2017,
    author = {S. Sasaki  and K. Hashimoto  and R. Kobayashi  and K. Itoh  and S. Iguchi  and Y. Nishio  and Y. Ikemoto  and T. Moriwaki  and N. Yoneyama  and M. Watanabe  and A. Ueda  and H. Mori  and K. Kobayashi  and R. Kumai  and Y. Murakami  and J. Müller  and T. Sasaki },
    title = {Crystallization and vitrification of electrons in a glass-forming charge liquid},
    journal = {Science},
    volume = {357},
    number = {6358},
    pages = {1381-1385},
    year = {2017},
    doi = {10.1126/science.aal3120},
    URL = {https://www.science.org/doi/abs/10.1126/science.aal3120},
}

@article{Scandurra2013,
    author = {Scandurra, G. and Giusi, G. and Ciofi, C.},
    title = "{Multichannel Amplifier Topologies for High-Sensitivity and Reduced Measurement Time in Voltage Noise Measurements}",
    journal = {IEEE Transactions on Instrumentation and Measurement},
    volume = {62},
    number = {5},
    pages = {1145},
    year = {2013},
    doi = {10.1109/TIM.2012.2236719},
}

@article{Thyzel2026SoftwarePackage,
  author       = {Thyzel, Tim},
  title        = {Software-defined lock-in amplifier (version 0.0.2)},
  year         = 2026,
  journal      = {Zenodo},
  doi          = {10.5281/zenodo.18631374},
  url          = {https://doi.org/10.5281/zenodo.18631374},
}

@article{Ryger2017,
    author = {Ryger, Ivan and Harber, Dave and Stephens, Michelle and White, Malcolm and Tomlin, Nathan and Spidell, Matthew and Lehman, John},
    title = {Noise characteristics of thermistors: Measurement methods and results of selected devices},
    journal = {Review of Scientific Instruments},
    volume = {88},
    number = {2},
    pages = {024707},
    year = {2017},
    month = {02},
    issn = {0034-6748},
    doi = {10.1063/1.4976029},
    url = {https://doi.org/10.1063/1.4976029},
}

@ARTICLE{Sampietro2000,
  author={Sampietro, N. and Accomando, G. and Fasoli, L.G. and Ferrari, G. and Gatti, E.C.},
  journal={IEEE Transactions on Instrumentation and Measurement},
  title={High sensitivity noise measurement with a correlation spectrum analyzer},
  year={2000},
  volume={49},
  number={4},
  pages={820-822},
  doi={10.1109/19.863931}
}

@Article{Schwalb2010,
AUTHOR = {Schwalb, Christian H. and Grimm, Christina and Baranowski, Markus and Sachser, Roland and Porrati, Fabrizio and Reith, Heiko and Das, Pintu and Müller, Jens and Völklein, Friedemann and Kaya, Alexander and Huth, Michael},
TITLE = {A Tunable Strain Sensor Using Nanogranular Metals},
JOURNAL = {Sensors},
VOLUME = {10},
YEAR = {2010},
NUMBER = {11},
PAGES = {9847--9856},
URL = {https://www.mdpi.com/1424-8220/10/11/9847},
DOI = {10.3390/s101109847}
}

@article{Achtenberg2021,
title = {Low-frequency noise measurements of IR photodetectors with voltage cross correlation system},
journal = {Measurement},
volume = {183},
pages = {109867},
year = {2021},
issn = {0263-2241},
doi = {https://doi.org/10.1016/j.measurement.2021.109867},
url = {https://www.sciencedirect.com/science/article/pii/S0263224121008095},
author = {Krzysztof Achtenberg and Janusz Mikołajczyk and Carmine Ciofi and Graziella Scandurra and Krystian Michalczewski and Zbigniew Bielecki},
}

@article{Scandurra2022,
    author = {Scandurra, G. and Ciofi, C. and Smulko, J. and Wen, H.},
    title = {A review of design approaches for the implementation of low-frequency noise measurement systems},
    journal = {Review of Scientific Instruments},
    volume = {93},
    number = {11},
    pages = {111101},
    year = {2022},
    month = {11},
    issn = {0034-6748},
    doi = {10.1063/5.0116589},
    url = {https://doi.org/10.1063/5.0116589},
}

\end{document}